\documentstyle[twoside,fleqn,espcrc2]{article}

\input{epsf}

\def\slash#1{#1 \hskip-0.50em /}
\def\Slash#1{#1 \hskip-0.65em /}
\def\Wc{W_{c}}
\def\WZdag{W\!Z^\dagger}

\newcommand{\AmS}{{\protect\the\textfont2
   A\kern-.1667em\lower.5ex\hbox{M}\kern-.125emS}}

\title{Heavy-to-light decays at large recoil:\\
Systematic treatment of short- and long-distance QCD effects\thanks{
Talk
presented at {\it QCD~02}\/,
  2-9~July~2002, Montpellier, France
(see {\tt home.cern.ch/\~{}tfeldman/qcd02\_talk.pdf}).
Based on work together with M.~Beneke, A.~Chapovsky,
           and M.~Diehl \cite{Beneke:2002ph}.}}

\author{Th.~Feldmann\address{Institut f\"ur Theoretische Physik~E,
RWTH Aachen, 52056 Aachen, Germany}%
}

\begin{document}

\begin{abstract}
Heavy quark decays into energetic, collinear quarks and gluons are
discussed within an effective theory that accomplishes the
factorization of soft and hard strong interaction effects.
We derive the relevant effective Lagrangian, and perform the
matching of the heavy quark current, including power corrections
of order $1/m$, where $m$ is the heavy quark mass.
We apply our framework to heavy-to-light form factors.
\hfil [{\tt hep-ph/0209239}]
\end{abstract}


\maketitle

\setcounter{footnote}{0}

\section{Introduction}

The phenomenological description of heavy quark decays requires to
minimize theoretical uncertainties related to strong interaction
effects.
One therefore tries to separate (factorize) calculable 
short-distance QCD contributions from 
long-distance physics related to
the non-perturbative binding of quarks and gluons. 
Short distance effects from integrating out
the weak gauge bosons and the top quark 
are included in the electroweak effective Hamiltonian
\cite{Buchalla:1995vs}.
Recently, the factorization of short-distance QCD dynamics
related to the heavy quark mass $m$ has been established 
\cite{Beneke:1999br,radiative1,radiative2,radiative2b} for 
exclusive heavy-to-light decays, in situations where
the recoil-energy to the light quark is large,
$E = {\cal O}(m)$. In the infinite mass limit, $m \to \infty$,
hadronic amplitudes can be expressed in terms of 
perturbatively calculable coefficient functions (hard-scattering
amplitudes) and hadronic transition form factors and light-cone
distribution amplitudes.

An elegant and efficient way to derive the factorization theorems
in cases like the one at hand,
is to match QCD  on an effective theory (``Soft-collinear effective
theory'', SCET) \cite{Bauer:2000yr}. The effective theory contains the 
field modes relevant to reproduce the infrared physics, while
the effect of hard modes is ``integrated out'' and encoded in
effective coupling constants (Wilson coefficients).
In the following, 
we will discuss the systematic derivation of power-corrections
within the effective theory approach, deriving both, 
the effective soft-collinear Lagrangian, and the
heavy quark current to order $1/m$ accuracy \cite{Beneke:2002ph}.
(Some aspects of power corrections in SCET
 have also been discussed in \cite{Chay:2002vy}.)

\section{Effective Lagrangian}

\begin{figure*}[bhpt]
\epsfxsize=0.65\textwidth
\fbox{\parbox{0.98\textwidth}{
\centerline{\epsffile{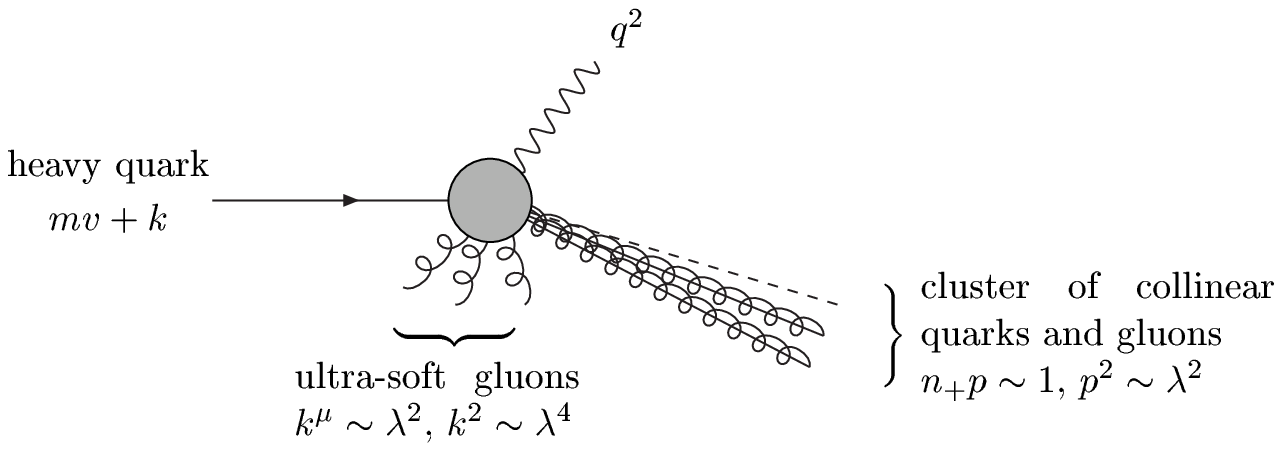}}
}}
\caption{Heavy quark decay into 
collinear and ultra-soft particles.}
\label{fig:setup}
\end{figure*}

In order to construct the effective Lagrangian, one first has
to identify the relevant degrees of freedom:
Nearly on-shell heavy quarks
are described within the well-known
heavy quark effective theory (HQET).
The different light quark and gluon modes 
are characterized by the scaling of momentum
components with respect to light-like vectors $n_+$ and $n_-$,
  \begin{equation}
    p^\mu \to (n_+p,p_\perp,n_- p) \ .
  \end{equation}
Collinear momenta scale as $p^\mu \sim m \, (1,\lambda,\lambda^2)$.
Ultra-soft particles have $p^\mu \sim m \, (\lambda^2,\lambda^2,\lambda^2)$.
Here the power-counting is expressed in terms of the small parameter
$\lambda \equiv (\Lambda_{\rm QCD}/m)^{1/2} \ll 1$
(the appropriate powers of the heavy quark mass $m$ will not be quoted
explicitly in the following).
Hard fields with $p^\mu \sim  (1,1,1)$ are to be integrated
out.\footnote{At one-loop order also soft particles with
   $p^\mu \sim (\lambda,\lambda,\lambda)$ may appear. 
   We will concentrate on power-corrections at tree-level, and
   the issue of soft modes will not be important
   for the further considerations.}
The kinematical situation to be described by the effective theory is
summarized in Fig.~\ref{fig:setup}.

At leading power, only collinear particles appear. To proceed,
it is convenient
to split the collinear quark spinor $\psi_c$ into its ``large'' and
``small'' components,
  \begin{eqnarray}
   \xi(x) &\equiv & \frac{\slash n_- \slash n_+}{4} \, \psi_{c}(x)
    \sim \lambda \ ,
 \cr
   \eta(x) &\equiv& \frac{\slash n_+\slash n_-}{4} \, \psi_{c}(x) \sim
  \lambda^2 \ ,
  \label{xieta}
    \end{eqnarray}
  where the scaling of the field components 
  with $\lambda$ follows from the  quark propagator in QCD.
  After integrating out the small spinor component $\eta(x)$, the
  leading-power collinear quark Lagrangian reads \cite{Bauer:2000yr}
 \begin{eqnarray}
  {\cal L}_{\xi}^{(0)} &=& 
   \bar \xi \, (in_-D_{c})\, \frac{\slash n_+}{2}   
      \, \xi
\label{Lcoll}
\\ &&
    {} +  \bar \xi \, i \Slash D_{\perp c}
     \, \frac{1}{(in_+D_{c}) +
       i\epsilon} \, i \Slash D_{\perp c} \, \frac{\slash n_+}{2}   
      \, \xi \ ,
\nonumber
  \end{eqnarray}
  where $D^\mu_c = \partial^\mu - i g A^\mu_c$ is the covariant
  derivative including collinear gluon fields.\footnote{The 
so-called large energy effective theory (LEET)
\cite{Dugan:1990de} does not contain collinear gluons and 
fails to reproduce the corresponding infrared properties of QCD.}
We note that 
the second term in (\ref{Lcoll}) is genuinely non-local and cannot be
  expanded into a series of local operators. Furthermore, the
  collinear Lagrangian lacks an intrinsic scale (the energy of
  collinear particles is not a Lorentz-invariant quantity). 
  As a consequence,
  the renormalization of the collinear effective Lagrangian
  is ``trivial'' if a Lorentz-invariant renormalization scheme
  like $\overline{\rm MS}$ is used, i.e.\ all UV-divergences can
  be absorbed into the running coupling constant.
  At sub-leading power in $\lambda$
  also ultra-soft quark and gluon fields 
  ($q_{\rm us}$, $A^\mu_{\rm us}$) have to be included. 
  Since ultra-soft fields have different characteristic wave-lengths
  than collinear fields, one has to perform a ``light-cone multipole
  expansion'' for vertices with collinear and ultra-soft particles. 
  This amounts to Taylor-expanding all ultra-soft fields
  around $x = (x_-,0_\perp,0)$, and applying leading-power equations of motion
  to obtain manifestly gauge-invariant results. (Notice that our
  formalism differs from \cite{Bauer:2000yr} where the large momentum
  components of collinear particles are extracted and collinear fields
  in the effective theory carry momentum labels.)
The induced power corrections to the collinear quark Lagrangian,
${\cal L}_{\rm \xi} = {\cal L}_{\rm \xi}^{(0)} + {\cal L}_{\rm
  \xi}^{(1)}+ {\cal L}_{\rm \xi}^{(2)} + \ldots $  read,\footnote{So
  far, the particular form of the Lagrangian (\ref{Lxi1}~--~\ref{Lxiq2}) has
been proven for the Abelian case, only.}
\begin{eqnarray}
  {\cal L}_{\xi}^{(1)}\!\! &\!\!=& 
 g  \,  \bar \xi  \,    \Wc  \,    (x_\perp^\mu    n_-^\nu    F_{\mu\nu}^{\rm us}) \, 
    \frac{\slash n_+}{2}   \,   \Wc^\dagger   \,  \xi \ ,
\label{Lxi1}
\\[0.1em]
 {\cal L}_{\xi}^{(2)} &=&
  \frac{g}{2}   \,   \bar \xi  \,    \Wc 
  \,  (n_-x)   \,   n_+^\mu n_-^\nu    F_{\mu\nu}^{\rm us}   \, 
   \frac{\slash  n_+}{2}   \,   \Wc^\dagger   \,   \xi
\cr
&& \!\!\!\!\!\!\! {} +  \frac{g}{2}   \,   \bar \xi    \,  \Wc 
   \,  x_\perp^\mu n_-^\nu \left[(x_\perp D_{\rm us}),   
    F_{\mu\nu}^{\rm us} \right]  
   \frac{\slash  n_+}{2}   \,   \Wc^\dagger   \,   \xi
\cr
&&\!\!\!\!\!\!\!  {} + \frac{g}{2}    \,  \bar \xi  
   \,   i \Slash D_{\perp c} \,    
 \frac{1}{i n_+ D_{c} \,}    \Wc   \,   x_\perp^\mu \gamma_\perp^\nu   
  F_{\mu\nu}^{\rm us}   \,   \Wc^\dagger  \, 
 \frac{\slash n_+}{2}    \,  \xi
\cr &&
\!\!\!\!\!\!\!  {} +  \frac{g}{2}    \,  \bar \xi  
  \,   \Wc   \,   x_\perp^\mu \gamma_\perp^\nu   
  F_{\mu\nu}^{\rm us}  \,     \Wc^\dagger   
 \frac{1}{i n_+ D_{c} \,}    i \Slash D_{\perp c} \,
 \frac{\slash n_+}{2}   \,   \xi \ . 
\cr && 
\end{eqnarray}
Beyond leading-power one also obtains
coupling terms between collinear and ultra-soft quarks,
\begin{eqnarray}
   {\cal L}^{(1)}_{\xi q_{\rm us}} &=& 
    \bar{\xi} \,  i\Slash{D}_{\perp c} \, W_{c} \,   q_{\rm us} + \bar
    q_{\rm us} \,  W_{c}^\dagger 
 i\Slash{D}_{\perp c} \,   \xi \ , 
\\[0.1em]
 {\cal L}^{(2)}_{\xi q_{\rm us}} &=&  
\bar{\xi}\,
 {i n_- D}\,    \Wc \,   \frac{\slash{n}_+}{2} \,  q_{\rm us} 
\cr
&&  \!\!\!\!\!\!\! {} +\bar{\xi}
   \,i \Slash{D}_{\perp  c} \,   \frac{1}{i n_+ D_{ c} \,}   
    i \Slash{D}_{\perp  c} \, W_{c} \,    
     \frac{\slash{n}_+}{2} \,  q_{\rm us}
\cr 
&&\!\!\!\!\!\!\! {} 
 +  \bar \xi  \,  i \Slash D_{\perp c} \,    W_{c}  \left[ { (x_\perp
  D_{\perp \rm us})}  \,   q_{\rm us}\right] +  \mbox{``h.c.''} \ .
\cr &&
\label{Lxiq2}
\end{eqnarray} 
Here $\Wc = P\exp [ig\int_{-\infty}^0 \!ds \,  n_+ A_c(x+s
    n_+)]$ is a collinear Wilson line, and
$D^\mu_{\rm us}$ and $F^{\mu\nu}_{\rm us}$ are the covariant
derivative and field strength tensor for ultra-soft gluon fields.
Note that all ultra-soft fields are understood to be taken at $x_-$.
Details of the calculation can be found in \cite{Beneke:2002ph}.

\section{Heavy quark current}
  
In the effective theory the heavy quark current
is expressed in terms of 
collinear and ultra-soft fields for
light quarks and gluons, and the HQET field
$h_v(x) = (1+\slash v)/2 \, e^{im \, v\cdot x} \, Q(x)$ for
heavy quarks moving with velocity $v^\mu$. In QCD 
the repeated emission of collinear (and ultra-soft) gluons 
puts the heavy quark off-shell, the intermediate
modes having large virtualities of order~1.
These hard modes have to be integrated out, which amounts
to resumming an infinite number of tree-level Feynman diagrams.
Alternatively, one solves the heavy quark Dirac
equation in the presence of collinear
and ultra-soft gluons, 
$
  (i\Slash D-m) \, Q(x) = 0 
$,
order by order in the expansion parameter
$\lambda$,
together with the appropriate boundary conditions 
for $A_{c}^\mu \to 0$. In both cases the solution
is constructed in terms of Wilson lines, 
\begin{eqnarray}
W(x) &=&
 P \, \exp \left[ig\int_{-\infty}^0 \!ds \,  n_+ A(x+s n_+)\right] \ ,
\cr
Z(x) &=&
 P\, \exp \left[ig\int_{-\infty}^0 \!ds \,  n_+ A_{\rm us}(x+s n_+)\right]
\cr &&
\label{Wilson}
\end{eqnarray}
where $W(x)$ contains both, collinear and ultra-soft gauge fields, 
while $Z(x)$ only involves ultra-soft gluons.
The Wilson lines (\ref{Wilson}) 
resum an infinite number of gluons, and their
properties under collinear and ultra-soft gauge transformations
are used to construct gauge-invariant operators in the effective theory.
Invariance under reparametrizations of the
heavy quark velocity \cite{Luke:1992cs}
can be made explicit by expressing the result in terms of 
the quantities
\begin{equation}
  Q_v(x) \equiv \left(1+ \frac{i\Slash D_{\rm us}}{m} + \ldots
\right) h_v(x) \ ,
\end{equation}
and
\begin{equation}
  {\cal V}^\mu \equiv v^\mu + \frac{ i D^\mu}{m}
\end{equation}
which transform homogeneously under $v \to v'$.
Including power corrections of order $\lambda^2$ the 
matching of the heavy quark field in QCD, $Q(x)$, and in
the effective theory, $Q_v(x)$, can be written as
\begin{eqnarray}
  && Q(x) = 
  e^{-imv\cdot x} \, \Big\{
  \WZdag \cr && \ {}- 
  \frac{1}{{\cal V}^2-1}\,
\Big[ \Slash{{\cal V}} \,\Slash{{\cal V}}\ \WZdag - \WZdag \, 
   \Slash{{\cal V}}_{\rm us} \,\Slash{{\cal V}}_{\rm us} \Big] 
  \Big\} Q_v(x)
\cr &&
\label{Qeff}
\end{eqnarray}
where the terms in the second line are power-suppressed with
respect to the leading term $ e^{-imv\cdot x} \, \WZdag  \, Q_v(x)
\simeq  e^{-imv\cdot x} \, \Wc  \, Q_v(x)$.
The expression (\ref{Qeff})
is also invariant under re\-parametri\-za\-tions 
of the reference vectors $n_\pm^\mu$. Again, the strict 
power-counting in $\lambda$ requires a multi-pole expansion
of the heavy quark current following from (\ref{xieta}) and
(\ref{Qeff}).
For details we refer to \cite{Beneke:2002ph}.

One-loop corrections to the leading-power 
heavy quark current in the
effective theory have been considered in \cite{Bauer:2000yr}.
The calculation is performed in momentum-space where 
renormalization is multiplicative. The result is easily translated
into our framework using convolution integrals in coordinate space
 \cite{Beneke:2002ph}.

\section{Heavy-to-light form factors}

The hadronic form factors for exclusive heavy-to-light decays 
are expressed in terms of 
matrix elements of the heavy quark current 
in the effective theory.
At leading power only the large spinor components $\xi$ and
$h_v$ are involved. As a consequence, the combined heavy mass/large
energy limit leads to new spin-/helicity-symmetries that 
reduce the number of independent form factors \cite{leet}.
For a $B$ meson decaying into a light pseudoscalar (vector) meson, one has 
one (two) universal form factor(s) instead of three (seven). 
A prominent phenomenological application
is the  lepton forward-backward asymmetry in $B \to K^*  \ell^+ \ell^-$
decay where, after taking into account next-to-leading order QCD
corrections, the correlation between the asymmetry zero in
the decay spectrum and the Wilson coefficient $C_9$ in the
electroweak Hamiltonian is predicted with about only 10\%
theoretical uncertainty \cite{radiative1}.

The inclusion of ${\cal O}(\lambda)$ corrections
to the heavy quark current increases the number of independent
form factors to two (five). On the ${\cal O}(\lambda^2)$ level
no form factor relations survive.
Each individual form factor $f_i(q^2)$ (where the index $i$ represents
a set of independent Dirac structures in the heavy quark current,
$\bar q \, \Gamma_i \, Q$) factorizes into soft
form factors $\xi_j^{(n)}(E)$ and hard-coefficient functions
$C_{ij}^{(n)}$ which are perturbatively calculable in the
effective theory. Schematically, for $B$ decays into light pseudoscalars
one has,
  \begin{eqnarray}
  && f_i(q^2) = C_i^{(0)} \, \xi^{(0)}(E)
\nonumber \\[0.15em]
  && \quad {}  +
           \sum_j \left( C_{ij}^{(1)} \, \xi_j^{(1)}(E)
          +  C_{ij}^{(2)} \, \xi_j^{(2)}(E) + \ldots \right)
\nonumber \\[0.1em]
  && \quad {}  + \mbox{hard spectator terms}
\label{factorize}
  \end{eqnarray}
and similarly for decays into transversely or longitudinally polarized
vector mesons. Here the index $j$ specifies an appropriate operator basis 
for heavy quark currents in the effective theory.
Little is known about the non-perturbative functions 
$\xi_j^{(1,2)}(E)$ that parameterize the power-corrections to
the form factor relations, and consequently phenomenological predictions
beyond leading-power remain model-dependent. However,
it is to be stressed that the effective theory framework enables us
to put these model estimates on a more solid theoretical basis.

The hard spectator terms, indicated in the last row
of (\ref{factorize}), have been calculated in QCD (using a
physical factorization prescription) to first
order in the strong coupling $\alpha_s$ \cite{radiative1}.
A treatment of hard spectator scattering within the 
effective theory framework is still missing.

\section{Conclusions}

The effective theory approach is an  elegant way to formulate the
factorization of short- and long-distance physics in processes with
different relevant energy scales.
The soft-collinear effective theory (SCET) describes the factorization
of collinear and ultra-soft field modes appearing, for instance, in heavy
quark decays into energetic light quarks. 
In the paper presented here \cite{Beneke:2002ph} we have shown how to
calculate power corrections of order $1/m$ 
to the effective SCET Lagrangian and to the heavy quark current
in a systematic way. 
Our results provide another step towards achieving
reliable theoretical predictions for $B$ meson decays,
that can be confronted to experimental data from $B$-factories
(see e.g.~\cite{BABAR,Belle}) in order to test the Standard Model 
against New Physics.

\end{document}